\documentclass[11pt]{article}
\def\fnote#1#2{\begingroup\def\thefootnote{#1}\footnote{#2}\addtocounter
{footnote}{-1}\endgroup}
\usepackage{graphicx}

\begin{document}

\hfill{UTTG-04-10}

\vspace{36pt}

\begin{center}
{\large {\bf { Ultraviolet Divergences in  Cosmological Correlations}}}

\vspace{36pt}
Steven Weinberg\fnote{*}{Electronic address:
weinberg@physics.utexas.edu}\\
{\em Theory Group, Department of Physics, University of
Texas\\
Austin, TX, 78712}

\vspace{30pt}

\noindent
{\bf Abstract}
\end{center}
\noindent
A method is developed for dealing with ultraviolet divergences in calculations of cosmological correlations, which does not depend on dimensional regularization.  An extended version of  the WKB approximation is used to analyze the divergences in these calculations, and these divergences are controlled by the introduction of  Pauli--Villars regulator fields.  This approach is illustrated in the theory of a scalar field with arbitrary self-interactions in a fixed flat-space Robertson--Walker metric with arbitrary scale factor $a(t)$.  Explicit formulas are given for the counterterms needed to cancel all dependence on the regulator properties, and an explicit prescription is given for calculating finite regulator-independent correlation functions.  The possibility of infrared divergences in this theory is briefly considered.

\vfill

\pagebreak

\begin{center}
{\bf I. INTRODUCTION}
\end{center}

Much effort has been expended in recent years in the calculation of quantum effects on cosmological correlations produced during inflation.  These calculations are complicated by the occurrence of ultraviolet divergences, which have typically been treated by the method of dimensional regularization.  Unfortunately, this method has several drawbacks.  It is difficult or impossible to employ dimensional regularization unless the analytic form of the integrand as a function of wave number is explicitly known, so calculations have generally relied on an assumption of slow roll inflation, or even strictly exponential inflation.  Also, even where an analytic form of the integrand is known, dimensional regularization can be tricky.  Senatore and Zaldarriaga[1] have shown that there are terms in correlation functions that were omitted in work by other authors[2],[3].  

This article will describe a method of dealing with ultraviolet divergences in cosmological correlations, without dimensional regularization.  For the purposes of regularization of infinities, we employ a generally covariant version of Pauli--Villars regularization[4].  In order to calculate the counterterms that are needed to cancel infinities when the regulator masses go to infinity, we introduce an extended version of the WKB approximation (keeping not only terms of leading order in wavelength), which works well even when the wave number dependence of the integrand is not explicitly known, and can therefore be applied for an arbitrary history of expansion during inflation.

This method is described here in a classic model, the fluctuations of a real scalar field in a fixed  general Robertson--Walker metric.  This is simple enough to illustrate the use of the method without the general idea being lost in the complications of quantum gravity, and yet sufficiently general so that we can see how to deal with an arbitrary expansion history.  
As we shall see, these methods yield a prescription for calculating correlation functions that  are not only free of ultraviolet divergences, but independent of the properties of the regulator fields.

\begin{center}
{\bf II. THE MODEL}
\end{center}

We consider the theory of a single real scalar field $\varphi(x)$ in a fixed metric $g_{\mu\nu}(x)$, with Lagrangian density
\begin{equation}
{\cal L}=\sqrt{-{\rm Det}g}\left[-\frac{1}{2}g^{\mu\nu}\partial_\mu\varphi \partial_\nu\varphi-V(\varphi)\right]\;,
\end{equation}
where $V(\varphi)$ is a general potential.  The modifications in this Lagrangian needed to introduce counterterms and regulator fields will be discussed in Sections III and IV, respectively.

This theory will be studied in the case of a general flat-space Robertson--Walker metric:
\begin{equation}
g_{00}=-1\;,~~~g_{0i}=0\;,~~~g_{ij}=a^2(t)\,\delta_{ij}\;,
\end{equation}
with $a(t)$ a fixed function (unrelated to $V(\varphi)$), which is arbitrary except that we assume that $a(t)$ increases monotonically from a value that vanishes for $t\rightarrow -\infty$.  The field equation is then
\begin{equation}
\ddot{\varphi}+3H\dot{\varphi}-a^{-2}\nabla^2\varphi+V'(\varphi)=0\;,
\end{equation}
where as usual $H\equiv \dot{a}/a$ is the expansion rate.
We define a fluctuation $\delta\varphi$ by writing
\begin{equation}
\varphi({\bf x},t)=\bar{\varphi}(t)+\delta\varphi({\bf x},t)\;,
\end{equation}
where $\bar{\varphi}(t)$ is a position-independent c-number solution of the field equation:
\begin{equation}
\ddot{\overline{\varphi}}+3H\dot{\overline{\varphi}}+V'(\overline{\varphi})=0\;.
\end{equation}

Our calculations will be done using an interaction picture, in which the time-dependence of $\delta\varphi$ is governed by the part of the Hamiltonian quadratic in $\delta\varphi$, so that $\delta\varphi$ satisfies a linear differential equation
\begin{equation}
\delta\ddot{\varphi}+3H\delta\dot{\varphi}-a^{-2}\nabla^2\delta\varphi+V''(\overline{\varphi})\delta\varphi=0\;.
\end{equation}
The commutation relations of $\delta\varphi$ are
\begin{equation}
[\delta\varphi({\bf x},t),\delta\dot{\varphi}({\bf y},t)]=ia^{-3}(t)\delta^3({\bf x}-{\bf y})\;,
\end{equation}
\begin{equation}
[\delta\varphi({\bf x},t),\delta\varphi({\bf y},t)]=[\delta\dot{\varphi}({\bf x},t),\delta\dot{\varphi}({\bf y},t)]=0\;.
\end{equation}
The fluctuation can therefore be expressed as
\begin{equation}
\delta\varphi({\bf x},t)=\int d^3q\;\left[e^{i{\bf q}\cdot{\bf x}}\alpha({\bf q})u_q(t)+
e^{-i{\bf q}\cdot{\bf x}}\alpha^\dagger({\bf q})u^*_q(t)\right]\;,
\end{equation}
where $\alpha({\bf q})$ is an operator satisfying the familiar commutation relations
\begin{equation}
[\alpha({\bf q}),\alpha^\dagger({\bf q}')]=\delta^3({\bf q}-{\bf q}')\;,~~[\alpha({\bf q}),\alpha({\bf q}')]=0\;,
\end{equation}
and $u_q(t)$ satisfies the differential equation
\begin{equation}
\ddot{u}_q+3H\dot{u}_q+a^{-2}q^2u_q+V''(\overline{\varphi})u_q=0
\end{equation}
and the initial condition, that for $t\rightarrow -\infty$,
\begin{equation}
u_q(t)\rightarrow \frac{1}{(2\pi)^{3/2}a(t)\sqrt{2q}}\exp\left[iq\int^{\cal T}_t\frac{dt'}{a(t')}\right]\\,
\end{equation}
where ${\cal T}$ is an arbitrary fixed time.
(The commutation relations (10) and the initial condition (12) ensure that the commutation relations (7) and (8) are satisfied for $t\rightarrow -\infty$.  The three commutators in these commutation relations satisfy coupled first-order differential equations in time, which with this initial condition imply that the commutation relations are satisfied for all times.)  

According to the ``in--in'' formalism[5], the vacuum expectation value of a product ${\cal O}_H(t)$ of Heisenberg picture fields and their derivatives, all at time $t$, is given by\footnote{It will be implicitly understood that the contours of integration over time are distorted at very early times to provide exponential convergence factors, as described in ref. [3].}
\begin{equation}
\left\langle {\cal O}_H(t) \right\rangle_{\rm VAC} =\left\langle \bar{T}\exp\left(i\int_{-\infty}^t H'_I(t')dt'\right){\cal O}_I(t)\;T
\exp\left(-i\int_{-\infty}^t H'_I(t')dt'\right)\right\rangle_0
\end{equation}
where $\langle\cdots\rangle_0$ denotes the expectation value in a bare vacuum state annihilated by $\alpha({\bf q})$; $T$ and $\bar{T}$ denote time-ordered and anti-time-ordered products; ${\cal O}_I(t)$ is the operator ${\cal O}(t)$ expressed in terms of interaction picture fluctuations; and $H'_I$ is the interaction Hamiltonian, the sum of terms in the Hamiltonian of third and higher order in the fluctuations, expressed in terms of the interaction-picture fluctuation $\delta\varphi$:
\begin{equation}
H'_I\equiv a^3\int d^3x\;\left[\frac{1}{6}V'''(\overline{\varphi})\delta\varphi^3+\frac{1}{24}V''''(\overline{\varphi})\delta\varphi^4+\dots
\right]
\end{equation}

We will evaluate Eq.~(13) as an expansion in the number of loops.  If we like, we can introduce a loop-counting parameter $g$ by writing $V(\varphi)=g^{-2}F(g\varphi)$, with $F(z)$ a $g$-independent function of $z$, so that the number of factors of $g$ in a diagram with $L$ loops and $E$ external scalar lines is 
\begin{equation}
\#=2L-2+E\;.
\end{equation}
Thus an expansion in the number of loops is the same as a series in powers of $g^2$.

\pagebreak

\begin{center}
{\bf III. ONE-LOOP COUNTERTERMS}
\end{center}

Infinities are encountered when calculating loop contributions to (13) in this model.  As in flat space, they can be canceled by introducing suitable counterterms into the Lagrangian.  (When regulator fields are introduced, the counterterms instead cancel dependence on the regulator properties.)  But the Lagrangian cannot know what metric  will be adopted, or the classical field $\overline{\varphi}$ around which the field $\varphi$ is to be expanded, so neither can the counterterms.  Thus we must return to the generally covariant form (1) of the Lagrangian in analyzing the possible counterterms that may be needed and employed.

The general one-loop one-particle-irreducible diagram consists of a loop into which are inserted a number of vertices, to each of which is attached any number of external lines.  An insertion with $N$ external lines is given by the $(N+2)$th derivative of $V(\varphi)$ with respect to $\varphi$ at $\varphi=\overline{\varphi}$, so 
 the counterterm in the Lagrangian can only be a function of $V''(\varphi)$, and of $g_{\mu\nu}$ and its derivatives.  Furthermore, the operators appearing in a counterterm needed to cancel infinities can only be of dimensionality (in powers of energy) four or less.  But $V''(\varphi)$ has dimensionality two, so the only generally covariant counterterm satisfying these conditions is of the form\footnote{This argument does not rule out an additional term proportional to $\sqrt{-{\rm Det}g}\,V''(\varphi)g^{\mu\nu}\partial_\mu\varphi\partial_\nu\varphi$, but one-loop diagrams do not generate ultraviolet divergent terms with spacetime derivatives acting on external line wave functions.}
\begin{equation}
{\cal L}^{1\;{\rm loop}}_{\infty}=\sqrt{-{\rm Det}g}\left[A\,V''(\varphi)+B[V''(\varphi)]^2+C\,R\,V''(\varphi)\right]\;,
\end{equation}
where $R$ is the usual scalar curvature, and $A$, $B$, and $C$ are constants that depend on the cutoff (that is, on the regulator masses), but not on the potential.  Dimensional analysis tells us that in the absence of regulator fields $A$ is quadratically divergent, while $B$ and $C$ are logarithmically divergent.  

If we now specialize to the Robertson--Walker metric (2), and write the scalar field as in (4), this counterterm becomes (aside from a c-number term)
\begin{eqnarray}
{\cal L}^{1\;{\rm loop}}_{\infty}&=&a^3\Bigg[A\,\Big(V'''(\overline{\varphi})\delta\varphi+\frac{1}{2}
V''''(\overline{\varphi})\delta\varphi^2+\dots\Big)\nonumber\\&&+ B\,\Big(2V''(\overline{\varphi})V'''(\overline{\varphi})\delta\varphi+[V'''^2(\overline{\varphi})+V''(\overline
{\varphi})V''''(\overline{\varphi})]\delta\varphi^2+\dots\Big)\nonumber\\&&-(6\dot{H}+12H^2)C\,\Big(V'''(\overline{\varphi})\delta\varphi + \frac{1}{2}
V''''(\overline{\varphi})\delta\varphi^2+\dots\Big)\Bigg]\;.
\end{eqnarray}
These terms are of one-loop order, and hence to that order are to be used only in the tree approximation, with a new term in the interaction Hamiltonian given by
\begin{equation}
\Delta H_I=-\int d^3 x\;{\cal L}^{1\;{\rm loop}}_{\infty}\;.
\end{equation}
The terms shown explicitly in Eq.~(17) are the only counterterms in Eq.~(18) that contribute in one-loop order to the one-point and two-point functions.

\begin{center}
{\bf IV. REGULATORS}
\end{center}

The counterterm (17) is certainly not the most general counterterm that would be consistent with the symmetries of the Robertson--Walker metric.  For instance, if we didn't know anything about general covariance, we would have no reason to expect that  $\dot{H}$ and $H^2$ should 
occur in the linear combination $R=-6\dot{H}-12H^2$.  In order to be sure that the divergences we encounter will be of a form that can be canceled by the counterterm (17),  although we do our calculations for the Robertson--Walker metric (2), we shall adopt a regulator scheme  derived from a generally covariant theory.

The usual approach to this problem is to use dimensional regularization, which we wish to avoid for reasons given in Section I.  There are other methods of regularization that have been extensively applied to the evaluation of expectation values of operators like the energy-momentum tensor in curved spacetimes[6] but not as far as I know to the calculation of cosmological correlations.  

One such method is covariant point-splitting[7].    This method is well suited to the calculation of expectation values of bilinear operators, where the ultraviolet divergence arises from the confluence of the arguments of the two operators.  Because it is a covariant method, it can be  implemented by a renormalization of the bilinear operator that respects its transformation and convergence properties.  It seems difficult to apply covariant point-splitting to the calculation of cosmological correlations, where one integrates over the separation of the spacetime arguments of the interaction Hamiltonian.'

There is another widely used method known as adiabatic regularization[8].  In this method, one subtracts from the integrand its asymptotic form for large wave numbers, as determined by an extended version of the WKB method.  Experience has shown that though not covariant, this method yields the same results for expectation values of bilinear operators as covariant point-splitting[9].  But adiabatic regularization affects the contribution of small as well as large internal wave numbers, so it seems unlikely that it can be applied to the calculation of cosmological correlations, where for some diagrams the contribution of small virtual wave numbers to correlation functions depends in a complicated way on external wave numbers, so that adiabatic regularization cannot be implemented by the introduction of generally covariant counterterms in the Lagrangian.

We will instead here employ a generally covariant version of Pauli--Villars regularization[4], which like covariant point splitting and adiabatic regularization has previously been applied to the calculation of expectation values.  For the theory studied here, the Lagrangian (1) is modified to read
\begin{eqnarray}
{\cal L}&=&\sqrt{-{\rm Det}g}\Bigg[-\frac{1}{2}g^{\mu\nu}\partial_\mu\varphi \partial_\nu\varphi-
\frac{1}{2}\sum_n Z_n \left(g^{\mu\nu}\partial_\mu\chi_n \partial_\nu\chi_n +M^2_n \chi_n^2\right)
\nonumber\\&& -V\left(\varphi+\sum_n\chi_n\right)\Bigg]
\;,
\end{eqnarray}
where $\chi_n$ are  regulator fields, and $Z_n$ and $M_n$ are real non-zero parameters.  In order to eliminate ultraviolet divergences up to some even order $D$, we must take the  $Z_n$ and regulator masses $M_n$ to satisfy 
\begin{equation}
\sum_n Z^{-1}_n=-1\;,~~\sum_n Z^{-1}_n M^2_n=0\;,~~\sum_n Z^{-1}_nM_n^4=0\;,~\dots\;,~\sum_n Z^{-1}_nM_n^D=0\;.
\end{equation}
For instance, if there were only logarithmic divergences then $D=0$, and we would only need one regulator field, with $Z_1=-1$.  In one-loop calculations the maximum degree of divergence is quadratic, i.e. $D=2$, and to satisfy the conditions (20) we need at least two regulator fields.  In our calculations we will not need to make a specific choice of the number of regulator fields, but  only assume that there are enough to satisfy Eq.~(20).

  The coefficients  $A$, $B$, and $C$ in the one-loop counterterm (17) will be given  values depending on the $Z_n$ and $M_n$, such that all expectation values (13) approach finite limits independent of the $Z_n$ and $M_n$, as the $M_n$ become infinite.  As we will see, this condition not only fixes the terms in $A$, $B$, and $C$ that are proportional to logarithms of regulator masses and the term in $A$ that is proportional to squares of regulator masses, but also the terms in $A$, $B$, and $C$ that depend on regulator masses only through their ratios, and hence that remain fixed as the regulator mass scale  goes to infinity.  The only terms in $A$, $B$, and $C$ that will not be fixed by this condition are finite terms independent of regulator properties, which of course represent the freedom we have to change the parameters in the potential or to add a non-minimal coupling of the scalar field to the curvature.

The regulator fields $\chi_n$ like the physical field $\varphi$ are written as classical fields plus fluctuations
\begin{equation}
\chi_n({\bf x},t)=\overline{\chi}_n(t)+\delta\chi_n({\bf x},t)\;.
\end{equation}
The classical fields satisfy the coupled field equations
\begin{eqnarray}
&&\ddot{\overline{\varphi}}+3H\dot{\overline{\varphi}}+V'\Big(\overline{\varphi}
+\sum_n\overline{\chi}_n\Big)=0 \\
&& \ddot{\overline{\chi}}_n+3H\dot{\overline{\chi}}_n+Z_n^{-1}V'\Big(\overline{\varphi}
+\sum_n\overline{\chi}_n\Big)+M_n^2\overline{\chi}_n=0 \;.
\end{eqnarray}
We assume throughout that the regulator masses $M_n$ are all much larger than $H(t')$ and $\left|V''\Big(\overline{\varphi}(t')\Big)\right|^{1/2}$ over the whole range from $t'\rightarrow -\infty$ to the time $t'=t$ at which the correlations are measured.  In consequence, the classical field equations (22) and (23) have a solution in which all the $\overline{\chi}_n$ are less than $\overline{\varphi}$ by factors of order $H^2/M_n^2$ and $|V''(\overline{\varphi})|/M_n^2$, and so may be neglected.  We adopt this solution for the classical fields.  In particular, the  field $\overline{\varphi}$ then satisfies the original classical field equation (5).

In dealing with internal lines, it is convenient to lump together the physical field fluctuation $\delta\varphi$ and the fluctuations $\delta\chi_n$ in the regulator fields, by introducing an index $N$ (and likewise $M$,  etc.) such that $\delta\chi_N$ is the physical field fluctuation $\delta\varphi$ for $N=0$ and is a  regulator field fluctuation for $N=n\geq 1$, both in the interaction  picture.  The general field fluctuations satisfy
the coupled field equations
\begin{equation}
\delta\ddot{\chi}_N+3H\delta\dot{\chi}_N-a^{-2}\nabla^2\delta\chi_N+M_N^2\delta\chi_N+Z_N^{-1}V''(\overline{\varphi})\sum_M\delta\chi_M=0\;,
\end{equation}
where $Z_0=1$ and $M_0=0$.
The commutation relations of the $\delta\chi$ are
\begin{equation}
[\delta\chi_N({\bf x},t),\delta\dot{\chi}_M({\bf y},t)]=ia^{-3}(t)\delta^3({\bf x}-{\bf y})Z_N^{-1}\delta_{NM}\;,
\end{equation}
\begin{equation}
[\delta\chi_N({\bf x},t),\delta\chi_M({\bf y},t)]=[\delta\dot{\chi}_N({\bf x},t),\delta\dot{\chi}_M({\bf y},t)]=0\;.
\end{equation}
The general fluctuation can therefore be expressed as
\begin{equation}
\delta\chi_N({\bf x},t)=\sum_M\int d^3q\;\left[e^{i{\bf q}\cdot{\bf x}}\alpha_M({\bf q})u_{Nq}^M(t)+
e^{-i{\bf q}\cdot{\bf x}}\alpha_M^\dagger({\bf q})u^{*M}_{Nq}(t)\right]\;,
\end{equation}
where $\alpha_N({\bf q})$ satisfy the  commutation relations
\begin{equation}
[\alpha_N({\bf q}),\alpha_M^\dagger({\bf q}')]=\delta^3({\bf q}-{\bf q}')Z_N^{-1}\delta_{NM}\;,~~[\alpha_N({\bf q}),\alpha_M({\bf q}')]=0\;,
\end{equation}
and the $u_{Nq}^M(t)$ are solutions of  Eq.~(24):
\begin{equation}
\ddot{u}^M_{Nq}+3H\dot{u}^M_{Nq}+a^{-2}q^2u^M_{Nq}+M_N^2u^M_{Nq}+Z_N^{-1}V''(\overline{\varphi})\sum_L u_{Lq}^M=0
\end{equation}
distinguished by the initial condition, that for $t\rightarrow -\infty$,
\begin{equation}
u_{Nq}^M(t)\rightarrow \frac{1}{(2\pi)^{3/2}a^{3/2}(t)\sqrt{2\kappa_{Nq}(t)}}\delta_N^M\exp\left[-i\int_{\cal T}^t\kappa_{Nq}(t')\,dt'\right]\;,
\end{equation}
where 
\begin{equation}
\kappa_{Nq}(t')\equiv \left(\frac{q^2}{a^2(t')}+M_N^2\right)^{1/2}\;.
\end{equation}
The $\alpha_N({\bf q})$ are all taken to annihilate the vacuum.
The two-point functions appearing in propagators are then given by
\begin{equation}
\left\langle\delta\chi_N({\bf x}_1,t_1)\delta\chi_M({\bf x}_2,t_2)\right\rangle_0=\sum_K\int d^3q\;e^{i{\bf q}\cdot({\bf x}_1-{\bf x}_2)}Z_K^{-1}u_{Nq}^K(t_1)\,u_{Mq}^{K*}(t_2)\;.
\end{equation}

In calculating one-loop graphs, we must integrate over one or more times $t_i$ associated with vertices, and over a single co-moving wave number ${\bf q}$.  There are two ranges of $q\equiv |{\bf q}|$ where the integrand is greatly simplified.  

In the first range, $q/a(t)$ (and hence all $q/a(t_i)$) is much greater than $H(t')$ and $\left|V''\Big(\overline{\varphi}(t')\Big)\right|^{1/2}$ for all $t'\leq t$, as well as much greater than the physical wave numbers associated with external lines, though  $q/a(t)$ is not necessarily greater than the regulator masses.  In this range, we can reliably evaluate the integrand in an extended version of the  WKB approximation, described in an Appendix.   Any term that would be convergent in the absence of cancelations among the physical and regulator fields makes a negligible contribution to the integral over this range.

In the second range, $q/a(t)$ is much less than the regulator masses,  though it is not necessarily less than   $H(t')$ or $\left|V''\Big(\overline{\varphi}(t')\Big)\right|^{1/2}$ or the physical wave numbers associated with external lines.  In this range, it is safe to ignore the regulator fields.  (We do not have to worry about the contribution of times $t'$ so much earlier than $t$ that $q/a(t')$ is of the order of the regulator masses, because this contribution is exponentially suppressed by the rapid oscillation of the integrand at these early times.)  

It is crucially important to our method of calculation that, because we assume that the regulator masses are much larger than  $H(t')$ and $\left|V''\Big(\overline{\varphi}(t')\Big)\right|^{1/2}$ and the physical wave numbers associated with external lines, these ranges of wave number {\em overlap}.  We can therefore separate the range of integration of co-moving wave number by introducing a quantity $Q$ in the overlap region, so that $Q/a(t)$ is much less than all regulator masses, {\em and} much greater than $H(t')$ and $\left|V''\Big(\overline{\varphi}(t')\Big)\right|^{1/2}$ and the physical wave numbers associated with external lines.  We can evaluate the integral over  $q\leq Q$ ignoring  the regulators, and over  $q\geq Q$ by  using the WKB approximation.   
No errors are introduced by this procedure in the final result, because we are taking the regulator masses to be arbitrarily large compared with $Q/a(t)$, which is taken to be arbitrarily large compared with $H(t')$ or $\left|V''\Big(\overline{\varphi}(t')\Big)\right|^{1/2}$ for $t'\leq t$ or the physical wave numbers associated with external lines, so terms proportional to quantities like $Q/M_na(t)$ or $Ha(t)/Q$ are entirely negligible.

It should be emphasized that $Q$ is neither an infrared nor an ultraviolet cutoff, but simply a more-or-less arbitrary point at which we choose to split the range of integration.  As long as $Q$ is chosen in the overlap of the two regions defined in the previous paragraphs, the sum of the integrals over $q\leq Q$ and $q\geq Q$ will automatically be independent of $Q$.                         

\begin{figure}
	\begin{center}
	\includegraphics{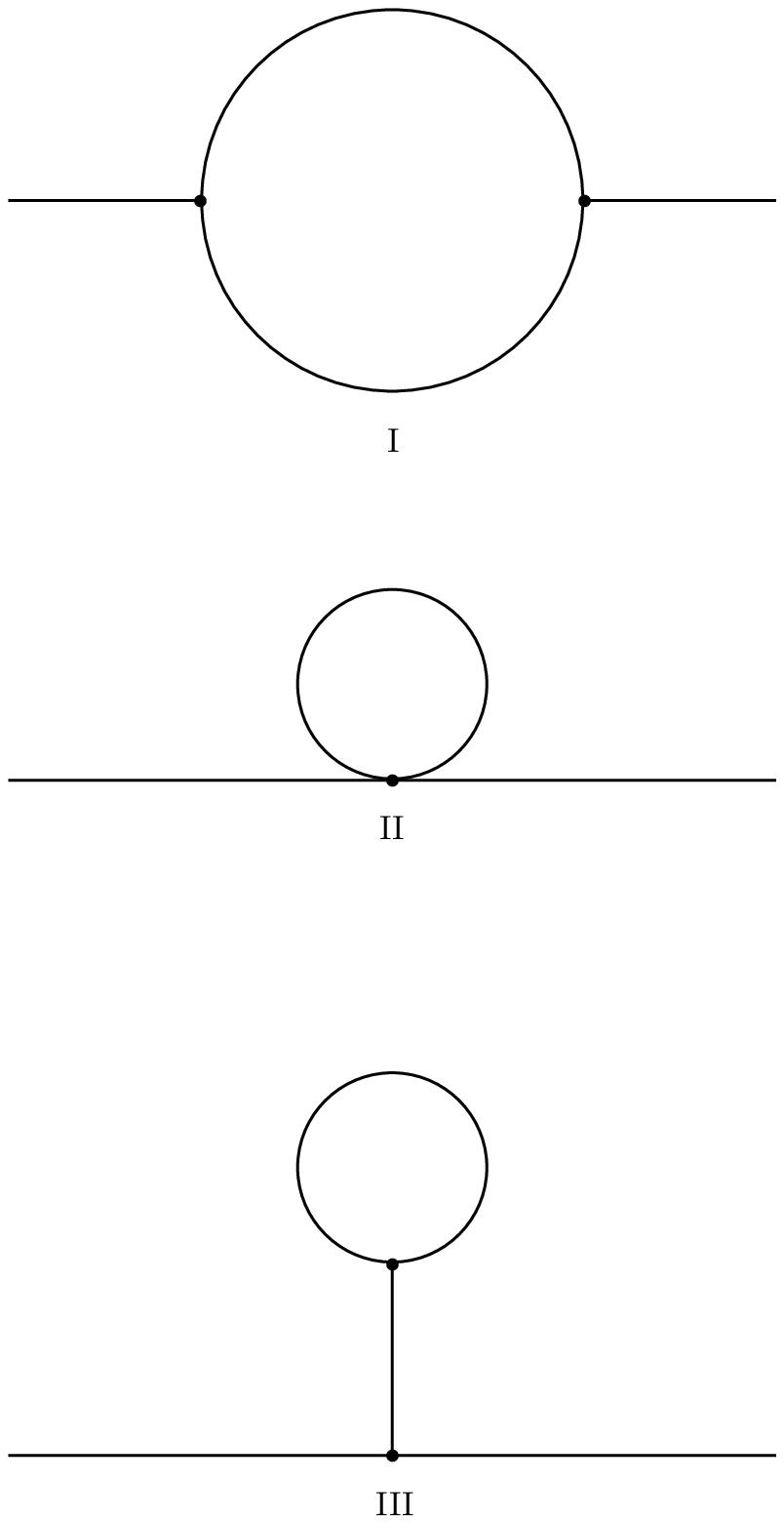}
	\end{center}
	\caption{Diagrams for the two-point function.}
\end{figure}

\begin{center}
{\bf V. THE TWO-POINT FUNCTION}
\end{center}

To demonstrate the use of the methods described in the previous section, and to evaluate the coefficients $A$, $B$, and $C$ in the counterterm (16), we will now calculate the one-loop corrections to the vacuum expectation value of the product $\delta\varphi_H({\bf y},t)\,\delta\varphi_H({\bf z},t)$ of Heisenberg picture fields.   
This can be written in terms of a Green's function $G_p(t)$, as
\begin{equation}
\left<\delta\varphi_H({\bf y},t)\,\delta\varphi_H({\bf z},t)\right>_{\rm VAC}=
\int d^3p\;\exp\Big(i{\bf p}\cdot{({\bf y}-\bf z})\Big)\,G_p(t)\;.
\end{equation}
Leaving aside vacuum fluctuations and counterterms, there are three one-loop diagrams, shown in Figure 1.  In this section we will consider only the one-particle-irreducible diagrams, I and II; these will suffice to allow us in Section VI  to fix the coefficients $A$, $B$, and $C$ in the counterterm (16).  Diagram III will be dealt with in Section VII.

\vspace{10pt}

\noindent
{\bf Diagram I}

By the usual rules of the ``in--in'' formalism, after integrating over spatial coordinates, the contribution of diagram I to the two-point function is
\begin{eqnarray}
G^I_p(t)&=&-2(2\pi)^6{\rm Re}\,\int_{-\infty}^t dt_1\;a^3(t_1)\,V'''\Big(\overline{\varphi}(t_1)\Big)
\,\int_{-\infty}^t dt_2\; a^3(t_2)\,V'''\Big(\overline{\varphi}(t_2)\Big)\nonumber\\
&&\times \sum_{KLMNM'N'}Z_K^{-1}Z_L^{-1}\int d^3q\;\nonumber\\&&\times \Bigg[\theta(t_1-t_2)u_p^2(t)u_p^*(t_1)u_p^*(t_2)
u_{Mq}^K(t_1)u_{M'q}^{K*}(t_2)u_{Nq'}^L(t_1)u_{N'q'}^{L*}(t_2)\nonumber\\&&
-\frac{1}{2}|u_p(t)|^2u_p^*(t_1)u_p(t_2)u_{Mq}^{K*}(t_1)u_{M'q}^{K}(t_2)u_{Nq'}^{L*}(t_1)u_{N'q'}^{L}(t_2)\Bigg]\;,
\end{eqnarray}
where $q\equiv |{\bf q}|$ and $q'\equiv |{\bf q}-{\bf p}|$.  The first term in the square brackets arises from diagrams in which the vertices come either both from the time-ordered product or both from the anti-time-ordered product in Eq.~(13), while the second term arises from diagrams in which one vertex comes from the time-ordered product and the other from the anti-time-ordered product.

As described at the end of the previous section, to calculate $G_p^I(t)$ we divide  the region of integration over 
$q\equiv |{\bf q}|$ into the ranges $q<Q$ and $q> Q$, where $Q$ is chosen so that $Q/a(t)$ is much less than all regulator masses but much greater than $p/a(t)$ and $H(t')$ and $|V''(\overline{\varphi}(t'))|^{1/2}$   for all $t'\leq t$.  

For $q<Q$, we can ignore the regulators, and set $K$, $L$, $M$, $N$, $M'$, $N'$ all equal to zero, with $u^0_{0q}$ just equal to the wave function $u_q$ in the absence of regulators.  This contribution takes the form
\begin{eqnarray}
G^{I,<Q}_p(t)&=&-2(2\pi)^6{\rm Re}\,\int_{-\infty}^t dt_1\;a^3(t_1)\,V'''\Big(\overline{\varphi}(t_1)\Big)
\,\int_{-\infty}^t dt_2\; a^3(t_2)\,V'''\Big(\overline{\varphi}(t_2)\Big)\nonumber\\
&&\times \Bigg[\theta(t_1-t_2)u_p^2(t)u_p^*(t_1)u_p^*(t_2)
\int_{q<Q} d^3q\;u_{q}(t_1)u_{q}^{*}(t_2)u_{q'}(t_1)u_{q'}^{*}(t_2)\nonumber\\&&
-\frac{1}{2}|u_p(t)|^2u_p^*(t_1)u_p(t_2)\int d^3q\;u_{q}^{*}(t_1)u_{q}(t_2)u_{q'}^*(t_1)u_{q'}(t_2)\Bigg]\;,
\end{eqnarray}
No limit has been put on the second integral over ${\bf q}$, because the oscillating exponentials in $u_q$ and $u_{q'}$ make this integral converge[3], so that the contribution of wave numbers with $q>Q$ is exponentially small. 

For $q\geq Q$, we can use the WKB approximation (30).  This contribution then takes the form
\begin{eqnarray}
G^{I,>Q}_p(t)&=&-2{\rm Re}\,\int_{-\infty}^t dt_1\;V'''\Big(\overline{\varphi}(t_1)\Big)
\,\int_{-\infty}^{t_1}dt_2\; V'''\Big(\overline{\varphi}(t_2)\Big)\,u_p^2(t)u_p^*(t_1)u_p^*(t_2)\nonumber\\
&&\times \sum_{KL}Z_K^{-1}Z_L^{-1}\int_{q>Q} \frac{d^3q}{4\sqrt{\kappa_{Kq}(t_1)\kappa_{Kq}(t_2)\kappa_{Lq}(t_1)\kappa_{Lq}(t_2)}}\;\nonumber\\&&\times 
\exp\left(-i\int_{t_2}^{t_1}[\kappa_{Kq}(t')+\kappa_{Lq}(t')]\,dt'\right)\;.
\end{eqnarray}
Note that we have dropped the distinction between $q'$ and $q$, because $p$ is negligible compared with $q$ for $q>Q$.  We have also dropped the contribution of the second term in Eq.~(34),  because this term converges for each $K$ and $L$, and so makes a negligible contribution to the integral over values $q>Q$. 

The contribution of values of $t_2$ at any fixed time less than $t_1$ is also negligible, because of the rapid oscillation of the final factor.  But there is an important contribution from values of $t_2$ that are so close to $t_1$ that $(t_1-t_2)Q/a(t_1)$ is not large.  This contribution can be evaluated by setting $t_2=t_1$ everywhere except in the range of integration in the exponential, so that
\begin{eqnarray}
G^{I,>Q}_p(t)&=&-2{\rm Re}\,\int_{-\infty}^t dt_1\;V'''\Big(\overline{\varphi}(t_1)\Big)
\,\int_{-\infty}^{t_1}dt_2\; V'''\Big(\overline{\varphi}(t_1)\Big)\,u_p^2(t)u_p^{*2}(t_1)\nonumber\\
&&\times \sum_{KL}Z_K^{-1}Z_L^{-1}\int_{q>Q} \frac{d^3q}{4\kappa_{Kq}(t_1)\kappa_{Lq}(t_1)}\;\nonumber\\&&\times \exp\left(-i(t_1-t_2)[\kappa_{Kq}(t_1)+\kappa_{Lq}(t_1)]\right)\nonumber\\&=&
-\int_{-\infty}^t dt_1\;V'''\Big(\overline{\varphi}(t_1)\Big)^2\,{\rm Im}\,\Big[u_p^2(t)u_p^{*2}(t_1)\Big]\nonumber\\
&&\times \sum_{KL}Z_K^{-1}Z_L^{-1}\int_{q>Q} \frac{d^3q}{2\kappa_{Kq}(t_1)\kappa_{Lq}(t_1)[\kappa_{Kq}(t_1)+\kappa_{Lq}(t_1)]}\;.
\end{eqnarray}
The integral over ${\bf q}$ converges because $\sum_K Z_K^{-1}=0$.  This integral receives contributions from terms where $\chi_K$ and $\chi_L$ are both regulator fields $\chi_m$ and $\chi_n$, or are a regulator field $\chi_n$ and a physical field $\chi_0=\varphi$, or are two physical fields.  Adding these contributions gives
\begin{eqnarray}
&&G^{I,>Q}_p(t)=
\pi\int_{-\infty}^t dt_1\;a^3(t_1)\,V'''\Big(\overline{\varphi}(t_1)\Big)^2\,{\rm Im}\,\Big[u_p^2(t)u_p^{*2}(t_1)\Big]\nonumber\\
&&~~~\times \left[\sum_{mn}Z_m^{-1}Z_n^{-1}\frac{M_n^2\ln M_n-M_m^2\ln M_m}{M_n^2-M_m^2}+2\sum_n Z_n^{-1}\ln M_n+\ln\left(\frac{Q}{a(t_1)}\right)\right]\;.~~~~~~
\end{eqnarray}
Note that, because $\sum_n Z_n^{-1}=-1$, this is independent of the units used to measure $Q$ and the regulator masses, as long as the same units are used in all logarithms.

\vspace{10pt}

\noindent
{\bf Diagram II}

By the usual rules of the ``in--in'' formalism, after integrating over spatial coordinates, the contribution of diagram II to the two-point function (33) is given by
\begin{eqnarray}
G^{II}_p(t)&=& (2\pi)^3 \int_{-\infty}^t dt_1\;a^3(t_1)\,V''''\Big(\overline{\varphi}(t_1)\Big)\,{\rm Im}\,\Big(u_p^2(t)u_p^{*2}(t_1)\Big)
\nonumber\\&& \times \sum_{KNN'}Z_K^{-1}\int d^3q\;u_{Nq}^K(t_1)u_{N'q}^{K*}(t_1)\;.
\end{eqnarray}
We again divide the range of integration over  
$q\equiv |{\bf q}|$ into the ranges $q<Q$ and $q\geq Q$, where $Q$ is chosen so that $Q/a(t)$ is much less than all regulator masses but much greater than $p/a(t)$ and $H(t')$ and $|V''(\varphi(t'))|^{1/2}$  for all $t'\leq t_1$. 

For $q<Q$ we can ignore the regulators, and set $K$, $N$, and $N'$ all equal to zero, with $u^0_{0q}$ just equal to the wave function $u_q$ in the absence of regulators.  This contribution takes the form
\begin{eqnarray}
G^{II,<Q}_p(t)&=& (2\pi)^3 \int_{-\infty}^t dt_1\; a^3(t_1)\,V''''\Big(\overline{\varphi}(t_1)\Big)\,{\rm Im}\,\Big(u_p^2(t)u_p^{*2}(t_1)\Big)
\nonumber\\&&\times \int_{q<Q} d^3q \;|u_{q}(t_1)|^2\;.
\end{eqnarray}

For $q>Q$ the individual terms in Eq.~(39) are quadratically divergent, so here we need an extended version of the WKB approximation (30), in which we keep terms in $u$ of order $\kappa^{-3/2}$ and $\kappa^{-5/2}$ as well as   $\kappa^{-1/2}$.  This is complicated by the presence of the potential term in Eq.~(29), which couples wave functions with different $\kappa$s.  We will deal with this by  considering  the potential term in Eq.~(29) as a perturbation.  Of course, $V''(\overline{\varphi})$ is {\em not} a perturbation; it is of zeroth order in the loop-counting parameter $g$ introduced at the end of Section II.  However, each insertion of $V''(\overline{\varphi})$ in the loop in Diagram II lowers its degree of divergence by two units, so the only terms we need to consider are those of zeroth and first order in $V''(\overline{\varphi})$, which in the absence of cancelations are quadratically and logarithmically divergent, respectively.  Terms of higher order in $V''(\overline{\varphi})$ are convergent even in the absence of cancelations, and are therefore negligible.

To evaluate the  terms in $G^{II,>Q}_p(t)$ of zeroth order in   $V''(\overline{\varphi})$, we note that in the absence of the potential, $u_N^K(t_1)$ is proportional to $\delta_{KN}$:
\begin{equation}
u_{Nq}^K(t_1) = \delta_{NK}u_{Nq}(t_1)\;,
\end{equation}
where
\begin{equation}
\ddot{u}_{Nq}+3H\dot{u}_{Nq}+(q^2/a^2)u_{Nq}+M_N^2 u_{Nq}=0\;.
\end{equation}
This contribution is
\begin{eqnarray}
G^{II, >Q, 0}_p(t)&=& (2\pi)^3 \int_{-\infty}^t dt_1\;a^3(t_1)\,V''''\Big(\overline{\varphi}(t_1)\Big)\,{\rm Im}\,\Big(u_p^2(t)u_p^{*2}(t_1)\Big)
\nonumber\\&& \times \sum_{N}Z_N^{-1}\int d^3q\;|u_{Nq}(t_1)|^2\;.
\end{eqnarray}
The integrand is given by an asymptotic expansion derived in the Appendix.  For both $q^2/a^2(t_1)$ and $M_N^2$ much greater than both $H^2(t_1)$ and $\dot{H}(t_1)$, we have
\begin{equation}
|u_{Nq}|^2\rightarrow \frac{1}{2\kappa_{Nq}a^3(2\pi)^3}\left[1+\frac{\dot{H}+2H^2}{2\kappa_{Nq}^2}+
\frac{(\dot{H}+3H^2)M_N^2}{4\kappa_{Nq}^4}-\frac{5H^2M_N^4}{8\kappa_{Nq}^6}\right]\;,
\end{equation}
where, as before, $\kappa_{Nq}^2(t_1)=\Big(q/a(t_1)\Big)^2+M_N^2$.  The integral over ${\bf q}$ converges because $\sum_N Z_N^{-1}=\sum_NZ_N^{-1}M_N^2=0$.  The sum over $N$ receives contributions from terms where $\chi_N$ is a regulator field $\chi_n$ or the physical field $\chi_0=\varphi$.  Adding these contributions gives
\begin{eqnarray}
G^{II, >Q, 0}_p(t)&=& \pi \int_{-\infty}^t dt_1\;a^3(t_1)\,V''''\Big(\overline{\varphi}(t_1)\Big)\,{\rm Im}\,\Big(u_p^2(t)u_p^{*2}(t_1)\Big)
\nonumber\\&& \times \Bigg[\sum_n Z_n^{-1}M_n^2\ln M_n+\Big(\dot{H}(t_1)+2H^2(t_1)\Big)\Big(\frac{5}{6}- \sum_n Z_n^{-1}\ln M_n\Big)\nonumber\\&&-\frac{Q^2}{a^2(t_1)}-\Big(\dot{H}(t_1)+2H^2(t_1)\Big)\ln\left(\frac{Q}{a(t_1)}\right)\Bigg]\;.
\end{eqnarray}

The regulator-dependent term arising from diagram II that are of first order in $V''(\overline{\varphi})$ can be calculated by applying  the rules of the ``in-in'' formalism a diagram like that of diagram II, but with a $V''$ insertion in the loop.  This gives
\begin{eqnarray}
G^{II, >Q, 1}_p(t)&=&-(2\pi)^6\int_{-\infty}^t dt_1\;a^3(t_1)\,V''''\Big(\overline{\varphi}(t_1)\Big)
\int_{-\infty}^t dt_2\; a^3(t_2)\,V''\Big(\overline{\varphi}(t_2)\Big)
\nonumber\\&&
\times\sum_{KLMNM'N'}Z_K^{-1}Z_L^{-1}\int_{q>Q}d^3q\;{\rm Re}\Bigg\{u_p^2(t)u_p^*(t_1)u_p^*(t_1)\nonumber\\&&
\times\left[\theta(t_1-t_2)u_{Mq}^K(t_1) u_{M'q}^{K*}(t_2)u_{Nq}^L(t_1) u_{N'q}^{L*}(t_2)+1\leftrightarrow 2\right]\Bigg\}.~~~~~~~
\end{eqnarray}
(This contribution is produced only by terms in which both interactions come from the time-ordered product in Eq.~(13), or both from the anti-time-ordered product.  As in the case of diagram I, the other terms make a negligible contribution to the part of the integral with $q>Q$.)  The individual terms in Eq.~(46) are only logarithmically divergent, so we can evaluate this using the leading term (30) in the WKB approximation.  Following the same limiting procedure as for diagram I, we find
\begin{eqnarray}
&&G^{II, >Q, 1}_p(t)=\pi\int_{-\infty}^t dt_1\;a^3(t_1)\,V''''\Big(\overline{\varphi}(t_1)\Big)V''\Big(\overline{\varphi}(t_1)\Big)\,{\rm Im}\,\Big[u_p^2(t)u_p^{*2}(t_1)\Big]\nonumber\\
&&~~~\times \left[\sum_{mn}Z_m^{-1}Z_n^{-1}\frac{M_n^2\ln M_n-M_m^2\ln M_m}{M_n^2-M_m^2}+2\sum_n Z_n^{-1}\ln M_n+\ln\left(\frac{Q}{a(t_1)}\right)\right]\;.~~~~~~
\end{eqnarray}

\vspace{10pt}

\noindent
{\bf Total 1PI Amplitude}

The complete contribution of the two one-particle irreducible diagrams is given by the sum of the terms (35), (38), (40), (45), and (47):
\begin{eqnarray}
&&G^{1PI}_p(t)=-2(2\pi)^6\int_{-\infty}^t dt_1\;a^3(t_1)V'''\Big(\overline{\varphi}(t_1)\Big)\int_{-\infty}^{t_1}dt_2\;a^3(t_2)V'''\Big(\overline{\varphi}(t_2)\Big)\nonumber\\&&
~~~\times{\rm Re}\left\{u_p^2(t)u_p^*(t_1)u_p^*(t_2)\int_{q<Q} d^3q\;u_q(t_1)u_q^*(t_2)u_{q'}(t_1)u_{q'}^*(t_2)\right\}\nonumber\\&&+(2\pi)^6\int^t_{-\infty} dt_1\;a^3(t_1)V'''\Big(\overline{\varphi}(t_1)\Big)\int_{-\infty}^t dt_2\;a^3(t_2)V'''\Big(\overline{\varphi}(t_2)\Big)\nonumber\\&&~~~\times |u_p(t)|^2{\rm Re}\left\{u_p^*(t_1)u_p(t_2)\int d^3q\;u_q^*(t_1)u_{q'}^*(t_1)u_q(t_2)u_{q'}(t_2)\right\}
\nonumber\\&& +(2\pi)^3\int_{-\infty}^t a^3(t_1)V''''\Big(\overline{\varphi}(t_1)\Big){\rm Im}\left\{u_p^2(t)u_p^*(t_1)\right\}\int_{q<Q}d^3q\;|u_q(t_1)|^2\nonumber\\&&
+\pi\int_{-\infty}^t dt_1\;a^3(t_1)\left[V'''\Big(\overline{\varphi}(t_1)\Big)^2+V''''\Big(\overline{\varphi}(t_1)\Big)V''\Big(\overline{\varphi}(t_1)\Big)\right]{\rm Im}\,\left\{u_p^2(t)u_p^{*2}(t_1)\right\}\nonumber\\&&~~\times \Bigg[\sum_{mn}Z_n^{-1}Z_m^{-1}\left(\frac{M_n^2\ln M_n-M_m^2\ln M_m}{M_n^2-M_m^2}\right)
+2\sum_n Z_n^{-1}\ln M_n+\ln\left(\frac{Q}{a(t_1)}\right)\Bigg]\nonumber\\&&
+\pi\int_{-\infty}^t dt_1\;a^3(t_1)V''''\Big(\overline{\varphi}(t_1)\Big){\rm Im}\,\left\{u_p^2(t)u_p^{*2}(t_1)\right\}\nonumber\\&&
~~~\times\Bigg[\sum_n Z_n^{-1}M_n^2\ln M_n-\frac{Q^2}{a^2(t_1)}\nonumber\\&&~~~~~~~~~~+\Big(\dot{H}(t_1)+2H^2(t_1)\Big)\left(\frac{5}{6}-\sum_n Z_n^{-1}\ln M_n-\ln\left(\frac{Q}{a(t_1)}\right)\right)\Bigg]\;.
\end{eqnarray}
To repeat, $q'\equiv |{\bf q}-{\bf p}|$, and $Q$ is any wave number for which $Q^2/a^2(t)$ is much larger than $H^2$ and $V''(\overline{\varphi})$ and $p^2/a^2(t)$ and much less than all regulator masses.  In this range, the $Q$-dependence of the first and third terms is canceled by the explicit $Q$-dependence of the fourth and fifth terms.

\begin{center}
{\bf VI.  CANCELING THE REGULATORS}
\end{center}

The terms in the counterterm (17) that are quadratic in the fluctuation make a contribution to the interaction-picture Hamiltonian of the form
\begin{equation}
\Delta H_I^{\rm quad}(t)=\frac{1}{2}{\cal G}(t)\int d^3x\;\delta\varphi^2({\bf x},t)\;,
\end{equation}
where 
\begin{equation}
{\cal G}=-a^3\left[AV''''(\overline{\varphi})+2B[V'''^2(\overline{\varphi})+V''(\overline{\varphi})V''''(\overline{\varphi})]-6C(\dot{H}+2H^2)V'''(\overline{\varphi})\right]
\end{equation}
According to the rules of the ``in-in'' formalism, this makes a contribution to the two-point function (33) given by 
\begin{equation}
\Delta G^{\rm 1PI}_p(t)=2(2\pi)^3\int_{-\infty}^t dt_1\;{\cal G}(t_1){\rm Im}\{u_p^2(t)u_p^{*2}(t_1)\}\;.
\end{equation}
Comparing Eqs.~(50) and (51) with (48), we see that in order to cancel the dependence of the one-particle irreducible two-point function on the regulator properties, we need
\begin{eqnarray}
A&=& \frac{1}{16\pi^2}\left[\sum_n Z_n^{-1}M_n^2\ln M_n+\mu_A^2\right]\\
B&=&\frac{1}{32\pi^2}\Bigg[\sum_{nm}Z_n^{-1}Z_m^{-1}\left(\frac{M_n^2\ln(M_n/\mu_B)-M_m^2\ln(M_m/\mu_B)}{M_n^2-M_m^2}\right)\nonumber\\&&+2\sum_n Z_n^{-1}\ln (M_n/\mu_B)\Bigg]\\C&=&-\frac{1}{96\pi^2}\left(\frac{5}{6}-\sum_n Z_n^{-1}\ln\left(\frac{M_n}{\mu_C}\right)\right)\;.
\end{eqnarray}
(The first term in Eq.~(52) does not depend on the units used for regulator masses in the logarithm, because $\sum_n Z_n^{-1}M_n^2=0$.)  Here $\mu_A$, $\mu_B$, and $\mu_C$ are unknown mass parameters.  The presence of these parameters should not be seen as a drawback of this method; they  reflect the real freedom we have to add finite regulator-independent terms to the original Lagrangian proportional to $V''(\varphi)$  or $V''^2(\varphi)$ or $R\,V''(\varphi)$. 

Adding Eqs.~(48) and (51) gives our final answer for the one-particle-irreducible part of the two-point function
\begin{eqnarray}
&&G^{1PI}_p(t)+\Delta G^{1PI}_p(t) =\Bigg[-2(2\pi)^6\int_{-\infty}^t dt_1\;a^3(t_1)V'''\Big(\overline{\varphi}(t_1)\Big)\int_{-\infty}^{t_1}dt_2\;a^3(t_2)V'''\Big(\overline{\varphi}(t_2)\Big)\nonumber\\&&
~~~\times{\rm Re}\left\{u_p^2(t)u_p^*(t_1)u_p^*(t_2)\int_{q<Q} d^3q\;u_q(t_1)u_q^*(t_2)u_{q'}(t_1)u_{q'}^*(t_2)\right\}\nonumber\\&&+ \pi \int_{-\infty}^t dt_1\; a^3(t_1)V'''\Big(\overline{\varphi}(t_1)\Big)^2{\rm Im}\{u_p^2(t)u_p^{*2}(t_1)\}\ln\left(\frac{Q}{a(t_1)\mu_B}\right)\Bigg]\nonumber\\&&+(2\pi)^6\int_{-\infty}^t dt_1\;a^3(t_1)V'''\Big(\overline{\varphi}(t_1)\Big)\int_{-\infty}^t dt_2\;a^3(t_2)V'''\Big(\overline{\varphi}(t_2)\Big)\nonumber\\&&~~~\times |u_p(t)|^2{\rm Re}\left\{u_p^*(t_1)u_p(t_2)\int d^3q\;u_q^*(t_1)u_{q'}^*(t_1)u_q(t_2)u_{q'}(t_2)\right\}
\nonumber\\&& +\Bigg[(2\pi)^3\int_{-\infty}^t dt_1\;a^3(t_1)V''''\Big(\overline{\varphi}(t_1)\Big){\rm Im}\left\{u_p^2(t)u_p^*(t_1)\right\}\int_{q<Q}d^3q\;|u_q(t_1)|^2\nonumber\\&&
+\pi \int_{-\infty}^t dt_1\; a^3(t_1)V''''\Big(\overline{\varphi}(t_1)\Big)\,{\rm Im}\{u_p^2(t)u_p^{*2}(t_1)\}\Bigg\{-\frac{Q^2}{a^2(t_1)}\nonumber\\&&~~+V''\Big(\overline{\varphi}(t_1)\Big)\ln\left(\frac{Q}{a(t_1)\mu_B}\right)-\Big(\dot{H}(t_1)+2H^2(t_1)\Big)\ln\left(\frac{Q}{a(t_1)\mu_C}\right)+\mu_A^2\Bigg\}\Bigg]\;.
\end{eqnarray}
For $Q^2/a^2(t)$ much larger than $H^2$, $\dot{H}$, $|V''(\overline{\varphi})|$, and $p^2/a^2(t)$, all $Q$ dependence cancels separately in the terms in square brackets on the first three lines and on the last three lines.  In this form, the two-point function (including also the one-particle-reducible contribution discussed in the following section) can be calculated even if all we have for the wave functions $u_q(t')$ is a numerical approximation.

\begin{center}
{\bf VII. ONE-PARTICLE-REDUCIBLE DIAGRAMS}
\end{center}

We now turn to the one-particle-reducible diagram III.   In this diagram the two external lines come together in a three-field vertex, with the third line terminating either in a three-field vertex to which is attached a scalar loop or a one-field vertex arising from the part of the one-loop counterterm (17) that is linear in $\delta\varphi$.  This part of the counterterm is
\begin{equation}
\Delta H_I^{\rm lin}(t)={\cal F}(t)\int d^3x\;\delta\varphi({\bf x},t)\;,
\end{equation}
with ${\cal F}(t)$ given by 
\begin{equation}
{\cal F}=-a^3\left[AV'''(\overline{\varphi})+2BV''(\overline{\varphi})V'''(\overline{\varphi})-C(6H^2+12\dot{H})V'''(\overline{\varphi})\right]\;.
\end{equation}
This diagram requires special treatment, because the line connecting the two vertices carries zero three-momentum.  For this reason, here we will delay integrating over the difference ${\bf x}$ of the spatial coordinate of the two vertices.    The full one-particle-reducible contribution to the two-point function (33) is then
\begin{eqnarray}
&& G_p^{1PR}(t)=2(2\pi)^3{\rm Re}\int_{-\infty}^t dt_1\;a^3(t_1)V'''\Big(\overline{\varphi}(t_1)\Big)
u_p^2(t_1)u_p^{*2}(t_1)\nonumber\\&&\times \int_{-\infty}^t dt_2\;{\cal I}(t_2)\int d^3x\;\Big[-\langle T\{\delta\varphi(0,t_1)\delta\varphi({\bf x},t_2)\}\rangle_0+\langle\delta\varphi({\bf x},t_1)\delta\varphi(0,t_2)\rangle_0\Big]\;,~~~
\end{eqnarray}
where
\begin{equation}
{\cal I}(t_2)\equiv \frac{1}{2}a^3(t_2)V'''\Big(\overline{\varphi}(t_2)\Big) \int d^3q\,\sum_{KNN'}u^K_{Nq}(t_2)u^{K*}_{N'q}(t_2)
+{\cal F}(t_2)\;.
\end{equation}
In the first term in the  square brackets in Eq.~(58), both vertices come from the time-ordered product in Eq.~(13), while in the second term, vertex 1 comes from the time-ordered product and vertex 2 from the anti-time-ordered product; in the complex conjugate time-ordered and anti-time-ordered products are interchanged.

There is no problem here with ultraviolet divergences coming from the integral over ${\bf q}$.  Following the same procedure as in our treatment of diagram II in the preceeding two sections, we have
\begin{eqnarray}
{\cal I}(t_2)&=&\frac{1}{2}a^3(t_2)V'''\Big(\overline{\varphi}(t_2)\Big)\Bigg[\int_{q<Q} d^3q |u_q(t_2)|^2+\frac{1}{8\pi^2}\Bigg(-\frac{Q^2}{a^2(t_2)}+V''\Big(\overline{\varphi}(t_2)\Big)\ln\left(\frac{Q}{a(t_2)\mu_B}\right)\nonumber\\&&
-\Big(\dot{H}(t_2)+2H^2(t_2)\Big)\ln\left(\frac{Q}{a(t_2)\mu_C}\right)+\mu_A^2\Bigg)\Bigg]\;,
\end{eqnarray}
where $Q$ is any wave number with $Q^2/a^2(t)$ much larger than  $\dot{H}(t')$ and $H^2(t')$ and $\left|V''\Big(\overline{\varphi}(t')\Big)\right|$ for all $t'\leq t$.  All dependence of $Q$ cancels in this limit.

But there is an apparent problem with infrared effects.   Eq.~(58) involves the integrals
$$ \int d^3x\;\langle T\left\{ \delta\varphi(0,t_1)\,\delta\varphi({\bf x},t_2)\right\}\rangle_0~~~{\rm and}~~~
\int d^3x \;\langle \delta\varphi({\bf x},t_2)\,\delta\varphi(0,t_1)\rangle_0\;.$$
When we use Eq.~(9) for the interaction-picture fields, the integrals over ${\bf x}$ pick out the value zero for 
the wave number $q$.  But the wave function $u_q(t)$ is not defined in the case $q=0$, because in this case there is of course no time early enough so that $q^2/a^2(t)$ is much larger than $H^2(t)$ and $|V''\Big(\overline{\varphi}(t)\Big)|$.  For the same reason, the argument for the Bunch--Davies condition $\alpha({\bf q})\Phi_0=0$ breaks down for ${\bf q}=0$.

Fortunately, we need the integrals over ${\bf x}$ only in the combination
\begin{eqnarray}
&& \int d^3x\;\left[-\langle T\left\{ \delta\varphi(0,t_1)\,\delta\varphi({\bf x},t_2)\right\}\rangle_0+\langle \delta\varphi({\bf x},t_2)\,\delta\varphi(0,t_1)\rangle_0\right]
\nonumber\\&&~~~~~=i\theta(t_1-t_2)\,G(t_1,t_2)
\end{eqnarray}
where 
\begin{equation}
G(t_1,t_2)\equiv i\int d^3x\;\left\langle \Big[\delta\varphi(0,t_1)\,,\,\delta\varphi({\bf x},t_2)\Big]\right\rangle_0\;.
\end{equation}
Despite the ambiguity in $u_0(t)$ and the inapplicability of the Bunch--Davies condition for ${\bf q}=0$, the function $G(t_1,t_2)$ is perfectly well-defined.  It is the solution of  the second-order differential equation
\begin{equation}
\left[\frac{d^2}{dt_1^2}+3H(t_1)\frac{d}{dt_1}+V''\Big(\overline{\varphi}(t_1)\Big)\right]\,G(t_1,t_2)=0\;,
\end{equation}
subject to initial conditions dictated by the commutation relations (7) and (8):
\begin{equation}
G(t_2,t_2)=0\;,
\end{equation}
\begin{equation}
\left[\frac{d}{dt_1}G(t_1,t_2)\right]_{t_1=t_2}=a^{-3}(t_2)\;.
\end{equation}
The only property of the vacuum state used here is that it has zero momentum and unit norm.
The general solution  is
\begin{equation}
G(t_1,t_2)=u(t_1)\,u(t_2)\int_{t_2}^{t_1}\frac{dt}{a^3(t)\,u^2(t)}\;,
\end{equation}
where $u(t)$ is any solution of the $q=0$ wave equation
\begin{equation}
\ddot{u}+3H\dot{u}+V''(\overline{\varphi})u=0\;,
\end{equation}
that does not vanish between $t_1$ and $t_2$.  
(For instance, for a general potential and a de Sitter metric, we can take $u=\dot{\overline{\varphi}}$,
which does not vanish in typical inflationary models.)
Putting this together, we have the one-particle-reducible contribution to the two-point function (33):
\begin{eqnarray}
&&G_p^{1PR}=
-2(2\pi)^3\int_{-\infty}^t dt_1\;a^3(t_1)\,V'''\Big(\overline{\varphi}(t_1)\Big)\,{\rm Im}\{u_p^2(t)u_k^{*2}(t_1)\}\nonumber\\&&~~~\times\int_{-\infty}^{t_1} dt_2 \,G(t_1,t_2)\,{\cal I}(t_2)\;.
\end{eqnarray}

\begin{center}
{\bf VIII. THE ONE-POINT FUNCTION}
\end{center}

In Section II we defined $\delta\varphi$ as the departure of the field $\varphi$ from its {\em classical} value $\overline{\varphi}$, not from its {\em mean} value, so we must expect $\delta\varphi$ to have a non-vanishing expectation value.  As we will see, this is closely related to quantities calculated in the previous section.

According to the general diagrammatic rules, the vacuum expectation value of the Heisenberg picture scalar field fluctuation in one-loop order is
\begin{equation}
\langle \delta\varphi_H({\bf y},t)\rangle^{\rm one\;loop}_{\rm VAC}=-i\int d^3x_1\int_{-\infty}^t dt_1\;
\langle \delta\varphi({\bf y},t)\,\delta\varphi({\bf x}_1,t_1)\rangle_0 
{\cal I}(t_1)+c.c.\;,
\end{equation} 
with ${\cal I}$ given by Eq.~(60) representing the insertion of a loop or a counterterm at the end of the single incoming line.  In the  term shown in Eq.~(69)  the single vertex comes from the time-ordered product in Eq.~(13); in its complex conjugate, the vertex comes from the anti-time-ordered product.  The two terms together involve the commutator of the field perturbations, so the one-point function may be written in terms of the function $G$ defined by Eq.~(62):
\begin{equation}
\langle \delta\varphi_H({\bf y},t)\rangle^{\rm one\;loop}_{\rm VAC}=-\int_{-\infty}^t dt_1\;
G(t,t_1)\, {\cal I}(t_1)\;.
\end{equation} 
We see now that the contribution (68) of the one-particle-reducible diagrams to the two-point function may be simply expressed in terms of the mean fluctuation:
\begin{eqnarray}
&&G^{1PR}_p(t)=\int_{-\infty}^t dt_1  \,a^3(t_1)\,V'''\Big(\overline{\varphi}(t_1)\Big)\nonumber\\&&~~~
\times\langle \delta\varphi_H(0,t_1)\rangle^{\rm one\;loop}_{\rm VAC}\;{\rm Im}\{u_k^2(t) u_k^{*2}(t_1)\}\;.
\end{eqnarray}
This is the same as would be given by adding  an interaction obtained by shifting $\delta\varphi$ by its expectation value:
\begin{equation}
\Delta H_I(t)=\frac{1}{2}a^3(t)\, V'''\Big(\overline{\varphi}(t)\Big)\,\langle \delta\varphi_H(0,t)\rangle^{\rm one\;loop}_{\rm VAC}\int d^3x \;\delta\varphi^2({\bf x},t)\;.
\end{equation}

\begin{center}
{\bf IX. INFRARED DIVERGENCES?}
\end{center}

Although the model treated in this paper is intended to provide an illustration of a method of dealing with ultraviolet divergences, it may be of some interest to look into the possible presence of infrared divergences in this model.  For any fixed co-moving wave number $q$, the evolution of the wave function $u_q(t)$ defined by Eqs.~(11) and (12) becomes $q$-independent once $q/a(t)$ drops below $H(t)$, so the behavior of the wave function for fixed $t$ and $q\rightarrow 0$ is determined by the behavior of $V''\Big(\overline{\varphi}(t')\Big)$ and $H(t')$ for $t'\rightarrow 0$.  We can distinguish two cases in which this problem is greatly simplified.

\vspace{10pt}

\noindent
{\bf Expansion-dominated}:\\
If $\left|V''\Big(\overline{\varphi}(t')\Big)\right|\ll H^2(t')$ for $t'\rightarrow 0$, then as long as this inequality is satisfied, 
we can drop the potential term in Eq.~(11), which then becomes the same as the differential equation for tensor fluctuations.  It is well known[10] in this case that if $\dot{H}(t')\rightarrow -\epsilon H^2(t')$ as $t'\rightarrow 0$, then the wave function $u_q(t_1)$ at a fixed time $t_1$ goes as $q^{-3/2-\epsilon}$ for  $q/a(t_1)\ll H(t_1)$.   This $q$-dependence is unaffected even if  
$H^2(t)$ drops below $\left|V''\Big(\overline{\varphi}(t)\Big)\right|$ at some time after $q/a$ drops below $H$, since the evolution of the wave function at such times is $q$-independent.  So (taking $\epsilon<1$) the integral over ${\bf q}$ of the product 
$u_q(t_1)u_q^*(t_2)$ in the propagator will be infrared divergent if and only if $\epsilon\geq 0$.  (We have been assuming  that as time passes fluctuations leave the horizon rather than entering it, so this discussion is limited to the case  $\epsilon<1$.  For the case $\epsilon\geq 1$, see ref. [11].)  There is no infrared divergence in the unlikely event that  the expansion rate increases at very early times.

\vspace{10pt}

\noindent
{\bf Potential-dominated}:\\
If $\left|V''\Big(\overline{\varphi}(t')\Big)\right|\gg H^2$ for $t'\rightarrow 0$, then as long as this inequality is satisfied, Eqs.~(11) and (12) have a WKB solution 
\begin{equation}
u_q(t')\simeq \frac{1}{(2\pi)^{3/2}a^{3/2}(t')\sqrt{2\omega(t')}}\exp\Big(i\int^{\cal T}_{t'}\omega(t'')\,dt''\Big)\;,
\end{equation}
where ${\cal T}$ is arbitrary, and 
\begin{equation}
\omega(t')\equiv \sqrt{\left(\frac{q}{a(t')}\right)^2+V''\Big(\overline{\varphi}(t')\Big)}\;.
\end{equation}
Once $q/a(t')$ falls below $|V''\Big(\overline{\varphi}(t')\Big)|$, the wave function $u_q(t')$ becomes independent of $q$, aside from a $q$-dependent phase that is independent of $t'$.  Later, $H^2(t')$ may or may not become comparable to or greater than $\left|V''\Big(\overline{\varphi}(t')\Big)\right|$, but this cannot affect the $q$-dependence of the wave function.  Therefore when the potential dominates at very early times, the product $u_q(t_1)u_q^*(t_2)$ in the propagator at fixed times $t_1$ and $t_2$ becomes $q$-independent for $q\rightarrow 0$, and there is no infrared divergence when we integrate the propagator over ${\bf q}$.

\begin{center}
{\bf X. FURTHER ISSUES}
\end{center}

The method described here can of course be applied in this model to all one-loop correlation functions.  The same counterterms, given by Eqs.~(16) or (17) and (52)--(54) will remove dependence on the regulator properties, because the only ultraviolet divergences in one-loop one-particle-irreducible diagrams occur in the one-point and two-point functions, which we have already discussed in Sections V through VIII.  The only ultraviolet divergences in higher correlation functions arise in diagrams in which trees are attached to loops at either one or two vertices, and the divergences in these loops are just those with which we have dealt.  Multi-loop graphs are more challenging.  

Beyond the simple model discussed here, of a scalar field in a fixed metric, there is the more realistic problem of scalar and tensor fluctuations in a theory of coupled scalar and gravitational fields.  This is  more complicated, because even in one-loop order there are quartic as well as quadratic and logarithmic ultraviolet divergences.  That alone should not prevent the  method described here from being applicable to realistic theories, at least for one-loop graphs, since divergences of any order can be eliminated by including enough regulator fields.

A more serious problem is the difficulty of introducing regulator fields for the graviton propagator.  (This problem is of course avoided in theories with large numbers of matter fields, where  matter loops dominate over graviton loops.)  If the only vertices that involve  gravitons have a single graviton line attached to matter lines, then we can introduce  regulators for the graviton propagator by coupling heavy tensor fields with suitable $Z$-factors to the energy-momentum tensor.  But it is not clear how to deal with graphs containing vertices to which are attached  two or more graviton lines.   

This raises the question whether Pauli--Villars regularization is really necessary.  The final results (55) and (68) for the one-particle irreducible and reducible parts of the two-point function could almost have been guessed without introducing regulator fields.  It would only be necessary to introduce an ultraviolet cut-off at a sufficiently large co-moving wave number $Q$, calculate the $Q$-dependence of the resulting two-point function by using the WKB methods described in this paper, and then introduce a counterterm of form (16), with $A$, $B$, and $C$ chosen as functions of $Q$ to cancel the $Q$-dependence found in this way.  (This is {\em not} the adiabatic regularization procedure mentioned in Section IV, even though both procedures use WKB methods, because with a cut-off at $Q$   only the part of the integrand for  internal wave numbers larger than $Q$ is affected.)  Of course, this procedure  leaves finite terms in $A$, $B$, and $C$ undetermined, but they are undetermined anyway, since they represent the real possibility of changing the original Lagrangian by adding corrections to the potential and adding a coupling of the scalar field to the spacetime curvature.  The cut-off introduced in this way would not respect general covariance, but apparently one would get the correct results (55) and (68) anyway.  

There is something mysterious about this.  The actual calculations in this paper were done for a fixed Robertson--Walker metric, Eq.~(2).  They would have been done in the same way by someone who had never heard of general covariance.  Yet the infinities turned out to depend on $H$ and $\dot{H}$ only in the combination $\dot{H}+2H^2$, proportional to the scalar spacetime curvature.  We can understand this for a generally covariant regularization procedure, like Pauli--Villars regularization, because in that case general covariance is broken only by the background, which presumably does not affect ultraviolet divergences.  But how do these calculations know that they are supposed to give infinities  that can be canceled by counterterms that are generally covariant, when we use a non-covariant cutoff on the internal wave number instead of introducing regulator fields?

\begin{center}
{\bf ACKNOWLEDGMENTS}
\end{center}

I am grateful for discussions with Joel Meyers, Emil Mottola, and Richard Woodard.  This material is based upon work supported by the National Science Foundation under Grant Numbers PHY-0969020 and PHY-0455649 and with support from The Robert A. Welch Foundation, Grant No. F-0014.

\renewcommand{\theequation}{A.\arabic{equation}}
\setcounter{equation}{0}

\begin{center}
{\bf APPENDIX: THE EXTENDED WKB APPROXIMATION}
\end{center}
We wish to find an asymptotic expression for the solution $u_q(t)$ of the differential equation
\begin{equation}
\ddot{u}_q(t)+3H(t)\dot{u}_q(t)+\Big(q^2/a^2(t)\Big)u_q(t)+M^2u_q(t)=0
\end{equation}
subject to the initial condition, that for $t\rightarrow 0$,
\begin{equation}
u_q(t)\rightarrow \frac{1}{(2\pi)^{3/2}a(t)\sqrt{2q}}\exp\left(iq\int_t^{\cal T} dt'/a(t')\right)\;.
\end{equation}
(The effects of the potential are treated separately in Section V.)  We are interested in the behavior of $u_q(t)$ at a fixed time $t$, when $q/a(t)$ is much larger than $H(t)$, but not necessarily greater than $M$.  

As an {\em ansatz}, we take
\begin{equation}
u_q(t)\rightarrow \frac{1}{(2\pi)^{3/2}a^{3/2}(t)\sqrt{2\kappa(t)}}\exp\left(i\int_t^{\cal T}\kappa(t') dt'\right)
\left[1+\frac{f(t)}{\kappa(t)}+\frac{g(t)}{\kappa^2(t)}+O(\kappa^{-3})\right]
\end{equation}
with $f$, $g$, etc. of zeroth order in  $q$ and $M$, and
\begin{equation}
\kappa(t)\equiv \sqrt{q^2/a^2(t)+M^2}\;.
\end{equation}
This clearly satisfies the initial condition (A.2).  The differential equation (A.1) is  satisfied by (A.3) to  order  $\kappa^{3/2}$ and $\kappa^{1/2}$, while the terms in (A.1) of order $\kappa^{-1/2}$ (counting $M$ as being the same order as $\kappa$) give
\begin{equation}
\frac{d}{dt}\left(\frac{f}{\kappa}\right)=\frac{i}{2\kappa}\left(\dot{H}+2H^2
+\frac{3H^2M^2}{2\kappa^2}-\frac{5M^4H^2}{4\kappa^4}+\frac{\dot{H}M^2}{2\kappa^2}\right)\;.
\end{equation}

The terms in (A.1) of order $\kappa^{-3/2}$ are more complicated, but fortunately we only need these terms in $|u_q|^2$, and for this purpose we can avoid having to work out these terms  by using the time-dependence of the Wronskian:
\begin{equation}
u_q^*\dot{u}_q-u_q\dot{u}^*_q\propto \frac{1}{a^3}\;.
\end{equation}
Using (A.3) gives
\begin{eqnarray}
&& 2(2\pi)^3a^3\Big(u_q^*\dot{u}_q-u_q\dot{u}^*_q\Big) =-2i-\frac{4i{\rm Re}\,f}{\kappa}
\nonumber\\&& +\frac{2i}{\kappa}\frac{d}{dt}\left(\frac{{\rm Im}f}{\kappa}\right)-2i\frac{|f|^2}{\kappa^2}
-\frac{4i{\rm Re}\,g}{\kappa^2}+O(\kappa^{-3})\;.
\end{eqnarray}

Now, Eq.~(A.5) shows that $d/dt(f/\kappa)$ is imaginary, so since $f(t)/\kappa(t)$ vanishes for $t\rightarrow 0$, $f(t)/\kappa(t)$ and hence $f(t)$ is imaginary for all $t$.  The first term on the right-hand side of Eq.~(A.7) is constant, and the second term vanishes, so the constancy of this quantity requires the vanishing of the terms of order $\kappa^{-2}$:
\begin{equation}
|f|^2+2{\rm Re}\,g=\kappa \frac{d}{dt}\left(\frac{{\rm Im}f}{\kappa}\right)
\end{equation}
But this is just what we need, for Eq.~(A.3) (with $f$ imaginary) gives
\begin{equation}
|u_q(t)|^2\rightarrow \frac{1}{2\kappa(t)(2\pi)^3a^3(t)}\left[1+\frac{|f(t)|^2+2{\rm Re}\,g(t)}{\kappa^2(t)}\right]\;.
\end{equation}
Together with Eqs.~(A.5) and (A.8), this gives the  result used in evaluating diagram II in Section V.
\begin{equation}
|u_{q}|^2\rightarrow \frac{1}{2\kappa a^3(2\pi)^3}\left[1+\frac{\dot{H}+2H^2}{2\kappa^2}+
\frac{(\dot{H}+3H^2)M^2}{4\kappa^4}-\frac{5H^2M^4}{8\kappa^6}\right]\;.
\end{equation}

\begin{center}
{\bf ----------}
\end{center}

\begin{enumerate}
\item L. Senatore and M. Zaldarriaga,   [0912.2734].
\item S. Weinberg, Phys. Rev. D {\bf 74}, 023508 (2006) [hep-th/0605244]; K. Chaicherdsakul, Phys. Rev. D {\bf 75}, 063522 (2007) [hep-th/0611352]; D. Seery, JCAP {\bf 0711}, 025 (2007) [0707.3377]; E. Dimastrogiovanni and N. Bartolo, JCAP {\bf 0811}, 016 (2008) [0807.2709]; P. Adshead, R. Easther, and E. A. Lim, Phys. Rev. D {\bf 79}, 063504 (2009) [0809.4008];  X. Gao and F. Xu, JCAP {\bf 0907}, 042 (2009) [0905.0405]; D. Campo, [0908.3642].
\item P. Adshead, R. Easther, and E. A. Lim, [0904.4207].
\item W. Pauli and F. Villars, Rev. Mod. Phys. {\bf 75}, 434 (1949).  Pauli-Villars regularization has been applied  to the problem of calculating the expectation value of the energy-momentum tensor in a  curved spacetime, by C. Bernard and A. Duncan, Ann. Phys. {\bf 107}, 201 (1977) and A. Vilenkin, Nuovo Cimento {\bf 44 A}, 441 (1977), but not as far as I know in the more complicated problem of calculating cosmological correlations.  The present work was done as a result of preparing a course on quantum field theory given in Spring 2010.
\item J. Schwinger, Proc. Nat. Acad. Sci. US {\bf 46}, 1401 (1960); J. Math. Phys. {\bf 2}, 407 (1961); K. T. Mahanthappa, Phys. Rev. {\bf 126}, 329 (1962); P. M. Bakshi and K. T. Mahanthappa, J. Math. Phys. {\bf 4}, 1, 12 (1963); L. V. Keldysh, Soviet Physics JETP {\bf 20}, 1018 (1965); D. Boyanovsky and H. J. de Vega, Ann. Phys. {\bf 307}, 335 (2003); B. DeWitt, {\em The Global Approach to Quantum Field Theory} (Clarendon Press, Oxford, 2003): Sec. 31.  For a review, with applications to cosmological correlations, see S. Weinberg, Phys. Rev. {\bf D72}, 043514 (2005) [hep-th/0506236].
\item For a general survey, see N. D. Birrell and P. C. W. Davies, {\em Quantum Fields in Curved Space} (Cambridge University Press, 1982).
\item S. M. Christensen, Phys. Rev. D 14, 2490 (1976).
\item L. Parker and S. A. Fulling, Phys. Rev. D 341 (1974); N. D. Birrell, Proc. Roy. Soc. (London) B361, 513 (1978); T. S. Bunch, J. Phys. A 13, 1297 (1980).
\item P. R. Anderson and L. Parker, Phys. Rev. D 36, 2963 (1987); S. Habib, C. MOlin-Paris and E. Mottola, Phys. Rev. D 61, 024010 (1999); P. Anderson, W. Eaker, S.Habib, C. Molina-Paris, and E.Mottola, Phys. Rev. D 62, 124019 (2000)
\item For a textbook treatment, see S. Weinberg, {\em Cosmology} (Oxford University Press, 2008), Sec. 10.3.
\item T. M. Jannsen, S. P. Miao, T. Prokpec, and R. P. Woodard, Class. Quant. Grav. 25, 245013 (2008) [0808.2449] 
\end{enumerate}

\end{document}